\documentclass[aps,pra,twocolumn,superscriptaddress,groupedaddress]{revtex4} 

\usepackage{graphicx}
\usepackage{amsmath}
\usepackage{amssymb}
\usepackage{mathtools}
\usepackage{dsfont}
\usepackage{bm}
\usepackage{color}
\usepackage{appendix}

\usepackage{amsthm}

\usepackage[usenames,dvipsnames]{xcolor}
\definecolor{nblue}{rgb}{0.3,0.3,1.0}

\def\n{n}
\def\QAOA{\mathrm{QAOA}}

\def\lf{\left\lfloor}   
\def\rf{\right\rfloor}
\newcommand{\z}{\sigma^z}
\newcommand{\x}{\sigma^x}
\newcommand{\y}{\sigma^y}

\newcommand{\bqa}{\begin{eqnarray}}
\newcommand{\eqa}{\end{eqnarray}}
\newcommand{\beq}{\begin{equation}}
\newcommand{\eeq}{\end{equation}}

\newcommand{\blue}{\textcolor{blue}}

\newcommand{\eq}{&=&}
\newcommand{\Otimes}{\bigotimes}
\def\bd#1{\boldsymbol{#1}}

\newcommand{\fid}{{\cal F}}
\newcommand{\bs}{\boldsymbol}
\newcommand{\nnm}{\nonumber\\}

\newcommand{\ea}{{\it et al.}}
\newcommand{\ges}{\geqslant}
\newcommand{\les}{\leqslant}
\newcommand{\maxcut}{MaxCut}
\newcommand{\Cmax}{C_{\text{max}}}

\newcommand{\pspin}{pseudospin}
\newcommand{\pspins}{pseudospins}

\usepackage{xy}
\xyoption{matrix}
\xyoption{frame}
\xyoption{arrow}
\xyoption{curve}
\usepackage{amsmath}
\entrymodifiers={}

\newcommand{\bra}[1]{{\left\langle{#1}\right\vert}}
\newcommand{\ket}[1]{{\left\vert{#1}\right\rangle}}

\newcommand{\ketbra}[2]{\ket{#1}\!\bra{#2}}

\newtheorem{theorem}{Theorem}

 \usepackage{amsmath}

\DeclareMathOperator{\tr}{Tr}

\usepackage{xcolor}
\definecolor{dark-red}{rgb}{0.4,0.15,0.15}
\definecolor{dark-blue}{rgb}{0.15,0.15,0.4}
\definecolor{medium-blue}{rgb}{0,0,0.5}
\usepackage{hyperref}
\usepackage[all]{hypcap}\usepackage{cleveref}\hypersetup{
    colorlinks, linkcolor={dark-red},
    citecolor={dark-blue}, urlcolor={medium-blue}
}

\def\QUAIL{
\affiliation{
Quantum Artificial Intelligence Laboratory (QuAIL), \\
NASA Ames Research Center, California 94035
}} 
\def\USRA{
\affiliation{
Universities Space Research Association, Columbia 21046 
}}
\def\SGT{
\affiliation{
Stinger Ghaffarian Technologies Inc., Maryland 20770
}}

\begin{document}

\title{Quantum Approximate Optimization Algorithm for MaxCut: A Fermionic View}
\author{Zhihui Wang}
\QUAIL 
\USRA
\author{Stuart Hadfield}
\affiliation{Department of Computer Science, Columbia University, New York 10027}
\author{Zhang Jiang}
\QUAIL
\SGT
\author{Eleanor G. Rieffel}
\QUAIL

\date{June 12, 2017; published: February 5, 2018; corrected: \today} 
\begin{abstract}
Farhi \ea\ recently proposed a class of quantum algorithms, 
the Quantum Approximate Optimization Algorithm (QAOA), for approximately
solving combinatorial optimization problems. A level-$p$ QAOA circuit 
consists of $p$ steps; in each step
a classical Hamiltonian, derived from the cost function, is applied followed
by a mixing Hamiltonian.
The $2p$ times for which these two Hamiltonians are applied 
are the parameters of the algorithm, 
which are to be optimized classically for the best performance.
As $p$ increases, parameter optimization
becomes inefficient due to the curse of dimensionality.
The success of the QAOA approach will depend, in part, on finding effective 
parameter-setting strategies.
Here, we analytically and numerically study parameter setting for QAOA applied
to \maxcut. For level-$1$ QAOA we derive an analytical expression for a
general graph.  
In principle, expressions for higher p could be derived, but the number of
terms quickly becomes prohibitive.  For a special case of \maxcut, the \lq\lq ring of
disagrees,\rq\rq\ or the one-dimensional antiferromagnetic ring, we provide an analysis for
arbitrarily high level. 
Using a fermionic representation, the evolution of the system under QAOA
translates into quantum control of an ensemble of independent spins.
This treatment enables us to obtain analytical expressions for the performance
of QAOA for any $p$. It also greatly simplifies numerical search for the
optimal values of the parameters.  By exploring symmetries, we identify a
lower-dimensional sub-manifold of 
interest; the search effort can be accordingly reduced. This analysis also 
explains an observed symmetry in the optimal parameter values.
Further, we numerically investigate the parameter landscape and show that it
is a simple one in the sense of having no local optima.\\

\noindent DOI: \href{https://doi.org/10.1103/PhysRevA.97.022304}{10.1103/PhysRevA.97.022304} 
\end{abstract}

 \maketitle

\section{Introduction}
Recently, Farhi \ea~\cite{Farhi2014}
proposed a new class of quantum algorithm, the Quantum Approximate
Optimization Algorithm (QAOA), to tackle
challenging approximate optimization problems on a gate model
quantum computer. 
In QAOA, the problem Hamiltonian, which encodes the
cost function of the optimization problem, and a mixing Hamiltonian are 
applied alternately. 
A handful of recent papers suggest the power of such circuits
\cite{Farhi2014b,Farhi2016,Shabani16,Jiang17}.
Once the problem and mixing Hamiltonians have been
chosen, the parameters of the algorithm are the times for which each
Hamiltonian is applied at each stage.
With an optimized time sequence for each piece, the optimal output
of the problem Hamiltonian is approximated.

The success of QAOA relies on being able to find a good time sequence.  
A level-$p$ algorithm has $2p$ parameters, the times (angles) for which the
problem Hamiltonian and the mixing Hamiltonian are applied at each iteration.
For QAOA of a fixed level, straight-forward sampling of search space was
proposed~\cite{Farhi2014}, but it is practical only for small $p$;
as the level increases the parameter optimization
becomes inefficient due to the curse of dimensionality.  
Elegant analytical tools designed for specific problem class can provide
parameter values for $p\gg 1$ that give near optimal performance, e.g., 
search an unstructured database~\cite{Jiang17}, but for a general problem, 
practical search strategies  are needed.
Here, we analytically and numerically study the
parameter setting problem, with a focus on the MaxCut
problem. We demonstrate how analyzing parameter symmetries and the 
landscape of the expectation value over the space of the parameter values
can aid in finding optimal parameter values.

In Ref.~\cite{Farhi2014}, Farhi \ea\ investigated MaxCut 
for specific (bounded-degree) graphs, and provided numerical results
for a special case, termed \emph{ring of disagrees}, which is a
one-dimensional chain of spin-1/2's with nearest-neighbored antiferromagnetic
couplings.  
We first extend the results of MaxCut to 
derive analytical expressions which can be solved to obtain the
optimal parameters for level-one QAOA for MaxCut on arbitrary graphs. 
Direct analysis through operator reduction quickly becomes cumbersome as the
level $p$ of the algorithm increases.
We focus on the ring of disagrees where we are able to advance the
analysis to arbitrary levels.  

Using a Fermionic
representation, we show that the evolution of the system under QAOA translates
into quantum optimal control of an ensemble of independent spins, 
significantly simplifying the analysis.  
In the new representation, the analytical expression for the expectation value
as a trigonometric polynomial of the parameters can be efficiently derived for
arbitrary level $p$. Furthermore, the reduction to independent spins
simplifies the numerical search greatly because 
evaluation involves only $\sim 2p$ matrix multiplications of 2-by-2
matrices and is linear in problem size, the number of spins in the original problem, $n$.
Further, by exploring symmetries, we identify a
lower-dimensional sub-manifold whose critical points are also critical
points in the full manifold. We numerically confirm for small $p$ that all 
optimal parameters lie in this sub-manifold.
The search effort can be accordingly reduced, and it also
explains an observed symmetry in the optimal parameter values.
Finally, a numerical investigation of the parameter landscape shows that it
is a simple one in the sense of having only global optima.

In Sec.~\ref{sec:recap} we give a recap of the QAOA algorithm, and a literature
review.  In Sec.~\ref{sec:maxcut} we introduce QAOA for MaxCut and present an
analytical expression for level-1 QAOA.
From Sec.~\ref{sec:ring} on, we focus on the anti-ferromagnetic chain (ring of
disagrees).  Sec.~\ref{sec:formulation} reviews the formulation of QAOA 
on this problem.
In Sec.~\ref{sec:fermionic}, we transform to a fermionic
representation and reduce the problem to the control of non-interacting spins.
Sec.~\ref{sec:saturation} provides analysis on the size-dependence of the
approximation ratio.
We analyze the symmetry in the system in Sec.~\ref{sec:symmetry}, and 
identify criticality-constrained manifolds where the global optima of the
parameters live.
In Sec.~\ref{sec:p1p2}, QAOA of level-1 and level-2 are illustrated;
In Sec.~\ref{sec:landscape}, we discuss the landscape topography of the search
space of the parameter values and its relation to known theory in quantum
control. 
Sec.~\ref{sec:conclusion} summarizes the main results and conclusions
of the paper.
 \section{Recap of the algorithm}
\label{sec:recap}
Given an objective function $C:\{0,1\}^\n \to \mathbf R$ to maximize, 
the aim of an
approximation algorithm is to find, upon specification of a desired 
approximation ratio $r^*$, a bit string $\mathbf x$ such that $C(\mathbf x)$
is within a factor of $r^*$ of the maximum:
\beq
\frac{C(\mathbf x)}{C_{max}} \ges r^*.
\eeq
An algorithm is an $r^*$-approximation algorithm for problem, if
for every instance of the problem, the algorithm finds a bit
string with cost function within $r^*$ of the maximum.
QAOA is a quantum approximate optimization algorithm that
iteratively alternates between applying a problem
Hamiltonian $H_C$ derived from the cost function and applying
a mixing Hamiltonian $H_B$, which in the standard case is
the transverse field $H_B = \sum_j \x_j$. For many problems, alternative mixing Hamiltonians that incorporate some 
problem constraints can reduce resource requirements and improve performance
over the standard setup~\cite{Hadfield17}.

From a classical cost function that is a polynomial in binary variables
$x_1, \dots, x_\n$, we can construct a Hamiltonian $H_C$ on
$\n$ qubits by first rewriting
the cost function in terms of variables $z_i\in\{-1,1\}$ where
$x_i = {(1 - z_i)}/{2}$ to obtain a polynomial
$f({\mathbf z}) 
= \sum_{{\cal C}\subset \{ 1,\ldots \n\}} \alpha_C \prod_{j\in {\cal C}}z_j$
and then replacing each occurance of $z_i$ with
the Pauli operator $\z_i$. Thus, $H_C$ is diagonal in the $\z$-basis and
takes the form
\bqa
H_C\eq \sum_{{\cal C} \subset \{ 1,\ldots \n\}} \alpha_{\cal C} \Otimes_{j\in {\cal C}} \z_j\;,
\eqa
where $\cal C$ is a subset of all qubits, 
and
$\alpha_{\cal C}$ is a real coefficient for the many-body coupling between
qubits in the subset $\cal C$.

We will use $\QAOA_p$ to refer to a level-$p$ QAOA circuit consisting 
of $p$ steps.
In each step, we first apply the problem Hamiltonian $H_C$, and then a mixing Hamiltonian $H_B$. 
Once the mixing Hamiltonian and the problem Hamiltonian have been
chosen, the parameters of a $\QAOA_p$ circuit are the $2p$ real
numbers $(\gamma_i, \beta_i)$, for $1 \les i \les p$, which determine
how long each operator is applied in iteration $i$:
\bqa
\label{eq:UC}
U_C(\gamma_i) \eq \exp[-i\gamma_i H_C] \\
\label{eq:UB}
U_B(\beta_i) \eq \exp[-i\beta_i H_B].
\eqa
Following Farhi \ea~\cite{Farhi2014b}, we refer to these times as
angles.
The standard initial state $\ket{\psi_0}$, a superposition of all classical
bit strings, is prepared as 
the ground state of $-H_B$ with 
density matrix 
\bqa
\label{eq:ini}
\rho_0 = \ket{\psi_0}\bra{\psi_0} = \Otimes_j \frac{1}{2}(\mathds{1}+\x_j).
\eqa
The circuit 
\begin{align}
\label{eq:U}
U=U_B(\beta_p)U_C(\gamma_p)\cdots U_B(\beta_2)U_C(\gamma_2)U_B(\beta_1)U_C(\gamma_1)
\end{align}
applied to the initial state creates a final state 
\beq
\ket{\boldsymbol{\gamma}, \boldsymbol{\beta}} = U\ket{\psi_0},
\eeq
for which the expectation value of $H_C$ is
\bqa
F( \boldsymbol{ \gamma}, \boldsymbol{  \beta})=\tr[ H_C U \rho_0 U^\dag ].
\label{eq:F}
\eqa

Let $F^*=F(\bs\gamma^*,\bs\beta^*)$ be the optimal value of $F$ over the value
range of the parameter set $\{(\bs\gamma, \bs\beta)\}$.  The approximation
ratio for the QAOA circuit with parameters $\{(\bs\gamma, \bs\beta)\}$ is
\bqa
 r\equiv{F}/\Cmax.
\eqa
The goal of the circuit $U$ is to drive the system into a quantum state which, 
upon measuring in the computational basis, 
yields with high probability a classical bit string that is $r^*$-approximately optimal.
This goal is achieved if the expectation value $F^*$ in the final state 
is $r^*$-approximately optimal, i.e., $r\ges r^*$,
and the distribution of bit strings from measuring 
this state in the computational basis is concentrated on bit strings
with costs close to this expectation value.

In Farhi \ea~\cite{Farhi2014b}, a $\QAOA_1$ algorithm beat the existing 
best approximation bound for efficient classical algorithms for
the problem E3Lin2, only to inspire a better classical
algorithm \cite{barak2015beating} that beats the approximation
ratio for the $\QAOA_1$ algorithm by a $\log$ factor. 
The performance of $\QAOA_p$ for E3Lin2 with $p > 1$ has yet to
be determined. Circuits with the above alternating structure have been
used for purposes other than approximate optimization. For example, such
QAOA circuits have also been applied for exact optimization
\cite{Jiang17,wecker2016training} and sampling \cite{Farhi2016}.
Wecker \ea~\cite{wecker2016training} explores learning parameters for
QAOA circuits on instances of MAX-2-SAT that result in high overlap
with the optimal state.
Jiang \ea~\cite{Jiang17} demonstrates that the class of QAOA circuits
is powerful enough to obtain the $\Theta(\sqrt{2^n})$ query complexity 
on Grover's problem, and also provides the 
first algorithm within the QAOA framework
to show a quantum advantage for a number of iterations $p$ in 
the intermediate range between $p=1$ and $p\to\infty$.
Farhi and Harrow~\cite{Farhi2016} proved that, 
under reasonable complexity assumptions, the output distribution of
even $\QAOA_1$ circuits cannot be efficiently sampled classically. 
QAOA circuits are therefore
among the most promising candidates for early demonstrations of
``quantum supremacy''~\cite{preskill_quantum_2012, boixo_characterizing_2016}.  
It remains an open question whether QAOA circuits provide a quantum
advantage for approximate optimization.

QAOA has close connection with the Variational Quantum Algorithm (VQA), 
in which classical optimization of parameters for a quantum evolution is
performed.  The result of evaluation
of the final state is fed back to the parameter optimization, forming a
closed-loop learning process.
Yang \ea~\cite{Shabani16} proved that for evolution under a Hamiltonian that
is the weighted sum of Hamiltonian terms, with the weights allowed to 
vary in time, the optimal control is bang-bang, i.e.
constant magnitude, of either the maximum or minimum allowed weight, 
for each of the terms in the Hamiltonian at any given time. 
Their work implies that QAOA circuits with the right parameters 
are optimal among Hamiltonians of the form $H(s) = \big(1-f(s)\big)H_B + f(s) H_C$,
where $f(s)$ is a real function in the range $[0,1]$.

The ultimate success of the QAOA approach will depend on finding effective
parameter-setting strategies.
For fixed $p$, the optimal parameters can be computed
in time polynomial in the number of qubits $n$~\cite{Farhi2014}.
With increasing $p$, however, exhaustive search of the
QAOA parameters becomes inefficient due to the
curse of dimensionality.
If we discretize so that each parameter can take on $m$ values,
exhaustive search of the optimum takes exponential
steps in $p$ as $m^{2p}$.
Here, we analytically and numerically study parameter setting for 
QAOA applied to \maxcut . 
 \section{{\bf QAOA}$_{\mathbf 1}$ for \maxcut }
\label{sec:maxcut}

In this section, we derive an analytical expression for the 
expecation value $F$ for 
$\QAOA_1$ for \maxcut\ on general graphs, 
furthering the analysis in \cite{Farhi2014}.
In principle, we could similarly derive expressions for 
higher $p$, but the workload quickly becomes prohibitive.

{\bf \maxcut\ Problem: }
Given a graph $G=(V,E)$ with $n=|V|$ vertices and $|E|$ edges, the
objective is to partition the graph vertices into two sets such that the
number of edges connecting vertices in different sets is maximized.

The cost function for \maxcut\ is
\begin{equation}
C=\frac{1}{2}\sum_{(i,j)\in E}({1-z_i z_j})\label{eq:QUBO}
\end{equation}
where $z_i$ and $z_j$ are binary variables associated to the vertices in $V$
which assume value +$1$ or -$1$ depending on which of the two partitions
defined by the cut are assigned.  
The Hamiltonian corresponding to this cost function is
\bqa
H_C = \sum_{\langle uv \rangle\in E} C_{uv},  \;\;\;\;\;\;\;\;\;\;\;\ C_{uv}= \frac{1}{2}(I-\z_u\z_v)\;.
\eqa
The expectation value of $H_C$ in QAOA decomposes as
\bqa
F(\bs\gamma,\bs\beta) \eq \sum_{\langle uv \rangle\in E} \langle {C_{uv}} \rangle
\eqa
where $\langle{C_{uv}}\rangle:= \tr[ C_{uv} U \rho_0 U^\dag ]$.
As $\langle C_{uv}\rangle \les 1$, the expected approximation ratio is lower bounded as
\bqa
r\ges \frac{F(\bs\gamma,\bs\beta)}{|E|}.
\eqa
\begin{theorem}
\label{th:1}
For QAOA with $p=1$, for each edge $\langle uv \rangle$,
\bqa
\langle {C_{uv}}\rangle \eq \frac12 + \frac14 (\sin 4\beta \sin \gamma) (\cos^{d_u} \gamma + \cos^{d_v} \gamma ) \nnm
&- & \frac14 (\sin^2 2\beta \cos^{{d_u}+{d_v}-2{\lambda_{uv}}}\gamma) (1-\cos^{\lambda_{uv}} 2\gamma), \nnm
\label{eq:th1}
\eqa
where ${d_u}+1$ and ${d_v}+1$ are the degrees of vertices $u$ and~$v$, respectively, and ${\lambda_{uv}}$ is the number of triangles in the graph containing edge $\langle uv \rangle\;.$
\end{theorem}

See Appendix~\ref{app:proof_maxcut} for a proof.  
The theorem implies that for $p=1$ the expectation value of any edge $\langle C_{uv} \rangle$ depends only on the parameters $({d_u},{d_v},{\lambda_{uv}})$. 
Then, the overall expectation value is
$F(\gamma,\beta)= \sum_{({d_1},{d_2},{\lambda})} \langle C_{uv} \rangle \chi ({d_1},{d_2},{\lambda})$, 
where the summation is taken over distinct subgraphs $({d_1},{d_2},{\lambda})$ 
and $\chi$ is the multiplicity of the subgraph, i.e. the number of times the
subgraph appears in $G$.  Thus, for an arbitrary graph the expectation value
$F(\gamma,\beta)$ may be efficiently computed classically, 
while to find an actual bit string realizing an approximate solution,
quantum computation resulting in the quantum state $U\ket{\psi_0}$
followed by measurement is required. 

\newtheorem{corollary}{Corollary}
\begin{corollary}
For a triangle-free $({d}+1)$-regular graph,  the expectation value of $\QAOA_1$ is
\begin{equation}    \label{eq:triangleFreeExpec}
F(\gamma,\beta) =  \frac{|E|}{2} \big(1+ \sin 4\beta \sin \gamma \cos^{d} \gamma\big)
\end{equation}
with maximum
\begin{equation} \label{eq:Fstar}
F^* =  \frac{|E|}{2} \Big(1+ \frac{1}{\sqrt {{d}+1}} \big(\frac{{d}}{{d}+1}\big)^\frac{{d}}{2}\Big) =: C^{\text{reg}}_{\max}({d})\;.
\end{equation}
For any such graph, one optimal pair of angles is $(\gamma,\beta)=( \arctan (1/\sqrt {d}),\pi/8)$.
\end{corollary}
Notice that $\big(\frac{{d}}{{d}+1}\big)^{{d}}>\frac{1}{e}$, 
the optimal approximation ratio is lower-bounded as
\bqa
r > \frac{1}{2}\Big( 1+\frac{1}{\sqrt e}\frac{1}{\sqrt{{d}+1}} \Big)
\eqa

\maxcut\ in the case of a regular graph of degree $2$, so the
graph is a ring, is termed the ring of disagrees in Ref.~\cite{Farhi2014}. 
In this case, for even $n$, the optimal partition is simply to include every other vertex into one set and the rest into the other set, hence $C_{\max}=n$.
Equation~\eqref{eq:Fstar} yields the approximation ratio $0.75$ at
$(\beta,\gamma)=(\pi/8,\pi/4)$, reproducing the results in~\cite{Farhi2014}.
For triangle-free 3-regular graph (${d}=2$), the ratio is $0.692$,
also in agreement with the results of Ref.~\cite{Farhi2014} for a general 3-regular graph.

Because $C^{\text{reg}}_{\max}({d})>|E|/2$ holds for all ${d}$, and
the values are concentrated around the expectation,
$\QAOA_1$ beats random guessing for arbitrary triangle-free regular graphs.
For an arbitrary triangle-free graph with maximum vertex degree ${d}+1$, 
applying the value to the right-hand side of Eq.~\eqref{eq:Fstar} gives a lower bound to $F^*$.

While it is straightforward to extend the analysis in the proof of
Theorem~\ref{th:1} to QAOA of higher levels, the number of terms quickly
becomes prohibitive for direct calculation; 
many more non-commuting terms coming from the $U_C$'s and $U_B$'s must be
retained and carried through the calculation. 
The expectation value of a given
edge will also depend on its local graph topology, which becomes difficult to succinctly characterize as $p$ increases.
(See Appendix \ref{sec:p2+} for the expression for the ring
of disagrees for $p = 2$.)

 \section{Analysis of the problem of ring of disagrees (Anti-ferromagnetic chain)}
\label{sec:ring}

We now study in detail QAOA for the ring of disagrees. 
We show that analysis can be done for $\QAOA_p$ for arbitrary level $p$,
extending numerical results for small $p$
given in Ref.~\cite{Farhi2014}.  

\subsection{Formulation of the problem}
\label{sec:formulation}
The Hamiltonian for the ring of disagrees with $n$ vertices 
i.e., a one-dimentional ring of spins of spin-${1}/{2}$, is
$\tilde H_C= \frac{1}{2} \sum_{j=1}^{n}(1- \z_j\z_{j+1})$
where $\z_{n+1}=\z_1$. 
For convenience, we later consider only even $n$, in which case the ground 
state of $\tilde H_C$ is trivial with every pair of neighboring spins 
aligned in anti-parallel fashion, corresponding to $\Cmax=n$. 
The approximation ratio is then $r=F^*/n$.

To simplify the derivation, and
also to conform to the convention in physics to minimize instead of maximize,
we drop the constant and rescale $\tilde H_C$ to be
\bqa
H_C = \sum_j \z_j\z_{j+1}\;,
\label{eq:H1}
\eqa
which defines the operator $U_C({\gamma})$ given in Eq.~\eqref{eq:UC}. 
The initial state of the system is prepared (Eq.~\eqref{eq:ini}), and 
the algorithm is specified by the QAOA circuit of Eq.~\eqref{eq:U}. 
Rewriting Eq.~(\ref{eq:F}) taking into account our simplification,
the approximation ratio for $F$ with parameters $(\beta, \gamma)$ is
\bqa
r=\frac{1}{2}\Big(1-\frac{F^*}{n}\Big)\;.
\eqa

The problem is now to determine parameters $(\beta, \gamma)$ that create a
quantum state that approximately 
minimizes the expectation value of $H_C$ (and thus maximizes $r$).
The relation between the angles and the expectation values used in the
remainder of this 
paper and the ones
in Ref.~\cite{Farhi2014} (notations with tilde) is $\gamma = -{\tilde
\gamma}/{2}$,
$\beta=\tilde\beta$
and
$\tilde F( \boldsymbol{\tilde \gamma}, \boldsymbol{\tilde  \beta})
=  \big(n- F( \boldsymbol{ \gamma}, \boldsymbol{  \beta})\big)/2$,
while the approximation ratio is the same.

 \subsection{Fermionic representation}
\label{sec:fermionic}
We show that using a fermionic representation, the parameter setting of QAOA
reduces to finding the optimal quantum control of an ensemble of independent spins
(spin-1/2). 

Since spin operators do not obey canonical commutation relations,
transforming them into bosonic or fermionic operators is often useful
for analysis. Such transformations enable the application of standard
techniques in condensed matter physics such as diagrammatic perturbation.
The algebra of the original spin operators must be preserved in the mappings.
The Jordan-Wigner transformation~\cite{lieb_two_1961, barouch_statistical_1970} maps the spin operators to fermions with a long-range phase factor.

We apply the Jordan-Wigner transformation~\cite{lieb_two_1961, barouch_statistical_1970},
\bqa
 a_j \eq  S^-_j e^{-i\phi_j} \\
 a_j^\dag \eq  S^+_j e^{i\phi_j}
\eqa
where $S^+_j=(\y_j+i\z_j)/2$, $S_j^-=(\y_j-i\z_j)/2$, and the phase factor $\phi_j=\pi \sum_{j'<j} (\x_{j'}+1)/2$ is long-ranged involving all operators for $j'<j$.
The new operators $a_j,~a_j^\dag$ can be verified to obey the fermion anticommutation relations, 
$\{a_j,a_{j'}^\dag\}=a_ja_{j'}^\dag+a_{j'}^\dag a_j=\delta_{j,j'}$, and
$\{a_j,a_{j'}\}=\{a_{j}^\dag,a_{j'}^\dag\}=0$.
The inverse transformation reads
\bqa
S_j^+ \eq  a_j^\dag e^{-i\phi_j}\\
S_j^- \eq  a_j e^{i\phi_j}\\
\x_j \eq 2 a^\dag_j a_j-1 
\eqa
and the phase factor in the fermionic representation is $\phi_j = \pi \sum_{j'<j} a^\dag_{j'} a_{j'}$.
The Jordan-Wigner transformation is a convenient tool for one-dimentional spin systems,
particularly for nearest-neighbored couplings because in products of the
neighboring spin operators like $S_j^+S_{j+1}^-$, the phase factors drop out,
leaving a concise expression with short-ranged coupling.

The Jordan-Wigner transformation for our problem works for both even and odd $n$. 
We will work on this general case in this section.
Applying the transformation to the problem and mixing Hamiltonians, we get
\bqa
H_B \eq \sum_{j=1}^{n} \big( 2a^\dag_j a_j -1\big)\\
H_C \eq \sum_{j=1}^{n-1} a^\dag_j a_{j+1}+a_ja_{j+1} - (a^\dag_Na_1+a_Na_1)G +\mathrm{h.c.}\;,\nnm
\eqa
where we introduce the gauge operator
$ G = \exp[i\pi\sum_{l=1}^n a^\dag_la_l] = (-1)^n \prod_{j=1}^n \x_j$,
a necessary treatment for periodic boundary conditions.
In the standard QAOA setting, 
the initial state is an eigenstate of $G$ with eigenvalue 1 for even $n$ and -1
for odd $n$.  The operator $G$ is a constant of motion since it commutes with
both $H_B$ and
$H_C$, so the value of $G$ remains constant throughout the evolution.
Therefore for even $n$, the sign of the $j=n$ term in $H_C$ is different 
from the others and requires a special treatment.

We further introduce a phase factor to unify the expression, $ b_j = a_j e^{-i j\pi /n}$.
The Hamiltonians then read
\bqa
H_B \eq \sum_{j=1}^n (2b_j^\dagger b_j-1)\\
H_C \eq e^{i \pi /n}\sum_{j=1}^{n}\Big( b_j^\dagger  b_{j+1} + e^{2i j\pi /n} b_j b_{j+1}\Big) +\mathrm{h.c.}
\eqa
Upon applying a Fourier transformation to $b_j$ (to $a_j$ for odd $n$), 
\begin{align}
 c_k = \frac{1}{\sqrt n}\sum_{j=1}^n e^{\omega j k}b_j\,,\quad \omega = 2i\pi /n\,,
\end{align}
the driver and the problem Hamiltonians in the momentum space take the form
\bqa
H_B \eq \sum_{k=0}^{n-1}  (2c_k^\dagger  c_k-1)\\
\label{eq:HCfermion}
\quad  H_C \eq 2\sum_{k=0}^{\lf{\frac{n-1}{2}}\rf} \cos\theta_k\, \big( c_k^\dagger c_{k} + c_{-k}^\dagger c_{-k}\big)  \nonumber\\
& &+i\sin\theta_{k} \big( c_k c_{-k} +  c_k^\dagger c_{-k}^\dagger \big)+H_{\text C,0} \eqa
where for even $n$
\bqa
\begin{cases}
H_{C,0} = 0 \cr
\theta_k = (2k+1)\pi /n \cr
c_{-k} \equiv c_{n-1-k}\;,
\end{cases}
\eqa
and for odd $n$
\bqa
\begin{cases}
H_{C,0} = -c_0^\dag c_0 \cr
\theta_k=2k\pi/n \cr
c_{-k}\equiv c_{n-k}\;.
\end{cases}
\eqa

Since in Eq.~\eqref{eq:HCfermion}, $c_k$ and
$c_k^\dagger$ are solely coupled to $c_{-k}$ and $c_{-k}^\dagger$, we only need
to solve a set of $2$-fermion problems.  Because both $H_B$ and $H_C$ 
preserve the
parity of the fermionic excitations, we need to consider only the ground state
and the double excited state of the two fermions. For each $k$, in this
two-dimensional subspace the driver and the problem Hamiltonians become
$2\z$ and $2\z\cos\theta_k + 2 \x\sin\theta_k$, respectively.

In summary, after transforming the problem to a fermionic representation, 
the original many-body Hamiltonian of a ring of $n$ spins reduces to an ensemble of $n/2$ non-interacting spins of spin-1/2, 
which we would refer to as \emph\pspins{} to distinguish them from spins in the
original problem:  
\bqa
H_B=\sum_{k=0}^{\lf \frac{n-1}{2}\rf} H_{B,k} \cr	H_C=\sum_{k=0}^{\lf \frac{n-1}{2}\rf} H_{C,k} 
\eqa
each term taking the form
\bqa
H_{B, k} \eq 2 \z \\
H_{C, k} \eq 2 \big(\cos\theta_k \z + \sin\theta_k\x \big) = 2\hat k \cdot \hat\sigma .
\eqa
where the unit vector $\hat k=(\sin\theta_k,0,\cos\theta_k)$.

The initial state for each \pspin{} is the ground state of $-H_{B,k}$, i.e.,  $\rho_0=\ketbra{1}{1}=({\mathds{1}}+\z)/2$ 
and the optimization reduces to minimize
\bqa
F( \boldsymbol{ \gamma}, \boldsymbol{  \beta}) \eq \sum_{k=0}^{\lf \frac{n-1}{2}\rf} F_k( \boldsymbol{ \gamma}, \boldsymbol{  \beta})\;,
\eqa
where
\bqa
F_k( \boldsymbol{ \gamma}, \boldsymbol{  \beta}) \eq \frac{1}{2}\big[ H_{C,k} U_k \z U_k^\dag \big] \\
\label{eq:Fs}
\eq \tr\big[ \hat k \cdot \hat\sigma  U_k \z U_k^\dag \big] \;.
\label{eq:Fs2}
\eqa
Hereafter, for notation simplicity, we drop the subscript for $U_k$ and use $U$ to refer to the  evolution operator for the single \pspin{}.
$
U = U_B(\beta_p)U_C(\gamma_p)\cdots U_B(\beta_1)U_C(\gamma_1)
$
now consists of only single-spin operators
\bqa
U_B(\beta_l) \eq \exp[-i2\beta_l \z] \\
U_C(\gamma_l) 
\eq \exp[-i2\gamma_l \hat k \cdot \hat \sigma ]
\eqa
for $l=1,2,\ldots,p$.
 \subsection{Size-dependence of the approximation ratio}
\label{sec:saturation}
For sufficiently large problem sizes, the approximation ratio of QAOA on the
problem of ring of disagrees of even $n$ is independent of the problem size.
This property has been shown in Ref.~\cite{Farhi2014} using an operator reduction argument.
The specific value of the approximation ratio for
$\QAOA_p$ was conjectured to be $(2p+1)/(2p+2)$ therein.
Here, we show that this size-independence feature comes naturally out of the
picture of single spin rotations.
Each $U_C(\gamma) = \cos(2\gamma) -i \sin(2\gamma) k\cdot \hat \sigma$ can
contribute a trigonometric function of $\theta_k$.  
Thus, $F_k$ takes a
form \bqa
\label{eq:FkTrig}
F_k \eq \sum_{\begin{array}{c}l,l'=0\\l+l'\les 2p+1\end{array}}^{2p+1} f_{l,l'}(\bs\gamma,\bs\beta) \sin^l \theta_k \cos^{l'} \theta_k\;,
\eqa
where $(f_{l,l'})$'s are real coefficients independent of $\theta_k$. 
Since each $\sin\theta_k$ accompanies one $\x$, using properties of
Pauli matrices, $\tr[\sigma_\alpha
\sigma_{\alpha'}]=2\delta_{\alpha,{\alpha'}}$, the coefficient
$f_{l,l'}(\bs\gamma,\bs\beta)$ is zero for odd $l$.  Recall that
$\theta_k=(2k-1)\pi/n$. 
When we consider $F$,
\begin{align}
\label{eq:FT}
F = \sum_{\begin{array}{c}l,l'=0\\l+l'\les 2p+1\end{array}}^{2p+1} \Big( f_{l,l'}(\bs\gamma,\bs\beta) \sum_{k=1}^{n/2} \sin^l \theta_k \cos^{l'} \theta_k \Big) \;,
\end{align}
for even $l$, we have
$
\sum_{k=1}^{n/2}\sin^l \theta_k \cos^{l'} \theta_k = 0
$
for odd $l'$.
Therefore, we need to keep only terms with even $l$ and even $l'$,
reducing  Eq.~(\ref{eq:FkTrig}), 
to a trigonometric polynomial of $2\theta_k$ of degree $p$, 
\bqa
F_k \eq \sum_{s=0}^p d_{2s} ( \boldsymbol{ \gamma}, \boldsymbol{  \beta}) \cos(2s\theta_k)\;,
\label{eq:FkFourier}
\eqa
where $d_{2s}(\bs\gamma,\bs\beta)$ is a coefficient independent of $k$.
See the analysis for $p=1$ and 2 in Sec.~\ref{sec:p1p2} for example.

Eq.~\eqref{eq:FkFourier} takes the form of the Fourier transformation of series
$d_{2s}$ with a cutoff at order $p$.  For any specific parameter values
$(\boldsymbol{ \gamma}, \boldsymbol{\beta})$, for $n\ges 2p+2$, 
the constant component
$d_0=\sum_k F_k/n$: 
\bqa
F\eq  \frac{n}{2}\cdot d_0 ( \boldsymbol{ \gamma}, \boldsymbol{  \beta})\;.
\eqa
Since the $n$-dependence of $F_k$ lies in $\theta_k$ and $d_0$ is
$\theta_k$-independent, the expectation value $F$, and furthermore the
approximation ratio of QAOA, is independent of $n$.
For an arbitrary level $p$, simplifying Eq.~\eqref{eq:Fs2} to get the specific
trigonometrical function form can be done easily.  Finding parameters
$( \boldsymbol{ \gamma}, \boldsymbol{  \beta})$ that optimize $F$
appears to be highly non-trivial.

 \subsection{Symmetry and criticality-constrained manifolds}
\label{sec:symmetry}
In this section, we show that, based on symmetries in the \pspin{} rotations,
we can identify sub-manifolds in the search space that admit extrema. In later
sections, we provide numerical evidence that the global minima always lie in
these sub-manifolds.
This evidence suggests that one can focus the search within the identified sub-manifolds and thus reduce the search effort.
 
\subsubsection{ Physics: rotations of the Bloch vectors}
\label{sec:rotation}
For each \pspin{}, Eq.~(\ref{eq:Fs}) can be expressed as
\bqa
F_k( \boldsymbol{ \gamma}, \boldsymbol{  \beta}) \eq 4\fid_k-2
\eqa
where \bqa
\fid_k &\equiv& \tr[\rho_{\hat k} U \rho_{z} U^\dag] 
\label{eq:Fid}
\eqa
and $\rho_{\hat k}=\frac{1}{2}(1+ \hat k\cdot \hat \sigma)$ and $\rho_z=\frac{1}{2} (1+\z)$.

On the Bloch sphere, $\rho_{\hat k}$ and $\rho_{\hat z}$ can be interpreted as
the density matrices for the Bloch vectors in the $\hat k$-direction and
$\hat z$-direction, respectively.  Equation~(\ref{eq:Fid}) represents a
single \pspin{}, initialized along $\hat z$-direction, then rotated about 
$\hat k$-axis for angle $4\gamma_1$, rotated about $\hat z$ for $4\beta_1$, ...,
rotated about $\hat k$ for $4\gamma_p$, rotated about $\hat z$ for $4\beta_p$,
and measured along $\hat k$. 
The fidelity ${\cal F}_k$ measures the overlap between the final state and the
state $\rho_{\hat k}$, whose Bloch vector is along direction $\hat k$.
Due to the periodicity in rotation, $F(4\bd\gamma+2\bd l\pi,4\bd\beta+2\bd l'\pi)=F(\bd\gamma,\bd\beta)\Rightarrow F(\bd\gamma+\bd l\cdot\pi/2,\bd\beta+\bd l'\cdot\pi/2)=F(\bd\gamma,\bd\beta)$, where $\bd l,~
\bd l'\in {\mathbb Z}^p$. Hence the search space can be limited to $\beta_k,\gamma_k\in[0,\pi/2]$ for $k=1,2,\ldots, p$. 

QAOA on the ring of disagrees thus corresponds to a physical picture
in optimal quantum control, albeit with a specialized set of constraints. 
For the final average over $k$ to get
$F$, we can think of the system as an ensemble of \pspins{}, each \pspin{} $k$
experiencing a constant magnetic field along $\hat k$, (the quantization
field), and the system controlled by applying a strong uniform magnetic field
along $\hat z$ in the ``bang-bang'' style. Specifically, when the field along
$\hat z$, $\vec B_z$, is on, the quantization field is negligible and all
\pspins{} are rotated about $\hat z$ by the same angle $4\gamma_p$; when $\vec B_z$
is paused,
each \pspin{} evolves freely, i.e., rotates about its own quantization axis
$\hat k$ to pick up an angle $4\beta_p$. After the whole control
sequence is applied, the overall magnetization along $\hat z$, $F=\sum_k\langle
\sigma^z_k \rangle$, is measured.  The goal of the quantum control is to design
a time sequence ($\bs \gamma,\bs \beta$) so that $F$ is minimized.

\subsubsection{Criticality-constrained sub-manifolds}
Since the trace operator preserves cycling, and the role of $\hat z$ 
and $\hat k$ in Eq.~\eqref{eq:Fid} are interchangeable, 
after initializing the \pspin{} along the $\hat k$-direction, 
the same result would be obtained by rotating
about $\hat z$-axis for angle $-4\beta_p$, 
rotating about $\hat k$ for $-4\gamma_p$,..., rotating about
$\hat z$ for $-4\beta_1$, rotating about $\hat k$ for $-4\gamma_1$, 
and measuring along $\hat z$.  

{\bf Manifold 1: }
Due to equivalence in the two views, it must hold that
\bqa
F_k(\bd \gamma,\bd\beta)=F_k(-\bd\beta',-\bd\gamma') 
\label{eq:symmetry}
\eqa
where 
\bqa
\bd\gamma \eq (\gamma_1,\gamma_2,\ldots , \gamma_{p-1},\gamma_p) \\
\bd\beta \eq (\beta_1,\beta_2,\ldots, \beta_{p-1},\beta_p) \\
\bd\gamma' \eq (\gamma_p,\gamma_{p-1},\ldots , \gamma_2, \gamma_1) \\
\bd\beta' \eq (\beta_p,\beta_{p-1},\ldots,  \beta_2, \beta_1) \;.
\eqa
This can be verified with the help of a $\pi$-rotation about the axis $\hat
z+\hat k$, see Appendix~\ref{sec:app1} for a proof. Consider the
manifold that satisfies 
\bqa
\gamma_i+\beta_{p+1-i}=0~\text{for}~ i=1,2,\ldots,p.
\label{eq:manifold1}
\eqa
It has a special property: the gradient of the function $F_k(
\boldsymbol{\gamma}, \boldsymbol{\beta})$ is constrained to lie tangent to the
manifold. Therefore, critical points in the manifold are critical points of the
whole function.

\begin{figure} [htbp]
\begin{center}  
\includegraphics[width=2in]{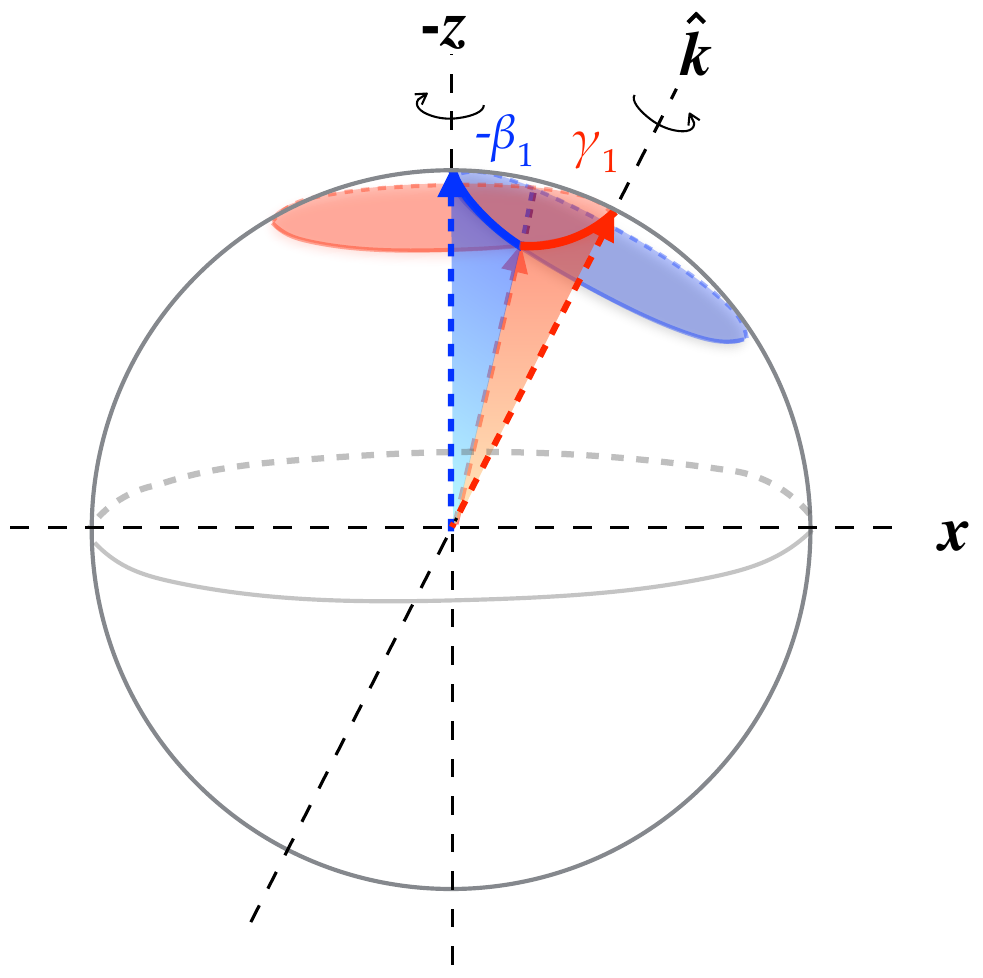}
\caption{
Schematic for the evolution trajectory of spin $k$ on the Bloch sphere under $\QAOA_1$ for arbitrary $\theta_k>\theta_1^*=2\pi/3$. 
The trajectory of $\QAOA_1$ is shown as arcs in solid lines.
\label{fig:rotation_p1}
}  
\end{center} 
\end{figure}

For $p=1$, the relation~\eqref{eq:manifold1} can also be proven to be a necessary condition for the global minima.
On the Bloch sphere for a \pspin, 
the rotation trajectory has to switch at the intersection of the two circles on the Bloch sphere 
which are perpendicular to one axis and passes through the vector end of the other \pspin{},
as illustrated in~Fig.~\ref{fig:rotation_p1}.  
Because the intersection lies in the plane spanned by $\hat y$ and the bisector
of $-\hat z$ and $\hat k$, it is obvious that $\gamma_1+\beta_1=0$, i.e., the
relation~\eqref{eq:manifold1} has to be observed.

{\bf Manifold 2: }
Equation~\eqref{eq:Fid} actually holds for $\rho_{\hat k}=\frac{1}{2}(1\pm \hat k\cdot \hat \sigma)$ and $\rho_z=\frac{1}{2} (1\pm \z)$.
The $+$ ($-$) sign correspond to the picture when the initial and 
final states are parallel (anti-parallel) with respect to the 
rotation axes, respectively.  Comparing these two pictures, since rotations 
by the same anble about any axis $-\hat v$ and $+\hat v$ by the same angle 
are inverses to each other, 
$\hat R_{(\hat v)}(\alpha)=\hat R^\dag_{(-\hat v)}(\alpha)$, 
it must hold that \bqa
F_k(\bd \gamma,\bd\beta)=F_k(\bd\beta',\bd\gamma')\;. 
\label{eq:symmetryB}
\eqa
Eq.~(\ref{eq:symmetryB}) defines another manifold
\bqa
\gamma_i-\beta_{p+1-i}=0~\text{for}~ i=1,2,\ldots,p
\label{eq:manifold2}
\eqa
with constrained gradient.

Eqs.~(\ref{eq:symmetry}) and (\ref{eq:symmetryB}) further indicate that $F_k$
and is an even function of the angle sequence: 
$F_k(\bd \gamma,\bd\beta)=F_k(-\bd \gamma,-\bd\beta),$
and accordingly so is $F$, 
\bqa
F(\bd \gamma,\bd\beta)=F(-\bd \gamma,-\bd\beta).
\label{eq:symmetry3}
\eqa

{\bf Global extrema lie in the submanifolds}
In our numerical search, the minima of $F$ were always contained in 
the manifold defined by Eq.~(\ref{eq:manifold1}) while the maxima 
of $F$ always lie in the manifold Eq.~(\ref{eq:manifold2}).

 \subsection{Illustration of {\bf QAOA}$_{\mathbf 1}$ and {\bf QAOA}$_{\mathbf 2}$ }
\label{sec:p1p2}

We use $\QAOA_1$ and $\QAOA_2$ to illustrate the results of symmetry and
size-dependence of the optimization discussed above.  Numerical results 
for higher levels ($p >2$) are shown in Appendix.~\ref{sec:p2+}

For $\QAOA_1$, the unitary evolution operator is
$
U= e^{-i2\beta_1 \z} e^{-i2\gamma_1 \hat k \cdot  \hat\sigma }\;.
$
Note that if a term $f(k)$ in $F_k$ satisfies $f(n/2+1-k)=-f(k)$, then $f(k)$  
would vanish in $F$ through the summation over $k$; and note the properties of Pauli matrices,
$\tr[\sigma_\alpha \sigma_{\alpha'}]=2\delta_{\alpha,{\alpha'}}$,
one comes to
\bqa
F \eq 2\sin(4\beta)\sin(4\gamma) \sum_k \sin^2\theta_k \\
\eq
\begin{cases}
n \sin(4\beta) \sin(4\gamma) & \text{for } n=2\\
\frac{n}{2} \sin(4\beta) \sin(4\gamma) & \text{for } n>2\;.\\
\end{cases}
\label{eq:p1}
\eqa
The optimal angles are $(\gamma_1^*,\beta_1^*)=\pi\cdot({3}/{8},{1}/{8})$ or $\pi\cdot(1/8,3/8)$. 

For $n=2$, the optimal angles correspond to $F^*=-n$ while for
larger problem size, $F^*=-n/2$.  This reflects the property that $\QAOA_p$ 
suffices to perfectly optimize the ring for $n\les 2p$ but for $n\ges 2p+2$
the optimization ratio is a fixed constant smaller than 1.

\begin{figure}  [htbp]
	\begin{center}  
		\includegraphics[width=\columnwidth]{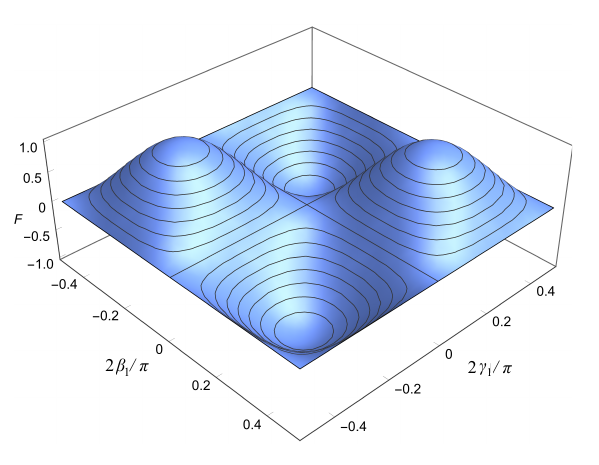}
		\caption{
			 $\QAOA_1$. The expectation value $F$ as a function of $\gamma_1$ and $\beta_1$.
						\label{fig:p1s}
			}  
	\end{center} 
\end{figure}

Eq.~(\ref{eq:p1}) is plotted in Fig.~\ref{fig:p1s}.  Along the symmetry line
$\beta_1+\gamma_1=0$,  the critical points are global minima and saddle points.
While along the symmetry line $\beta_1-\gamma_1=0$,  the critical points are
global maxima and saddle points.

 For level-2, the evolution operator reads
\bqa
U\eq e^{-i2\beta_2 \z} e^{-i2\gamma_2 \hat\sigma \cdot \hat k} e^{-i2\beta_1 \z} e^{-i2\gamma_1 \hat\sigma \cdot \hat k}\;.
\eqa
The expectation value $F$ as a trigonometric function of $(\bs \gamma, \bs \beta)$ is shown in Appendix~\ref{sec:p2+}.
Numerically found optimal angles are
$(\gamma_1^*,\beta_1^*,\gamma_2^*,\beta_2^*)=\pi\cdot(0.3956,~0.1978,~  0.3022,~0.1044)$
or $\pi\cdot(0.2052,~0.1026,~0.3974,~0.2948)$.
In both optimal angle sets, $4(\gamma_1^*+\beta_2^*)$ and $4(\gamma_2^*+\beta_1^*)$ are integer multipliers of $2\pi$, thus both optima lie in the manifold defined by Eq.~(\ref{eq:manifold1}).

\begin{figure} [htbp]
	\begin{center}  
		\includegraphics[width=\columnwidth]{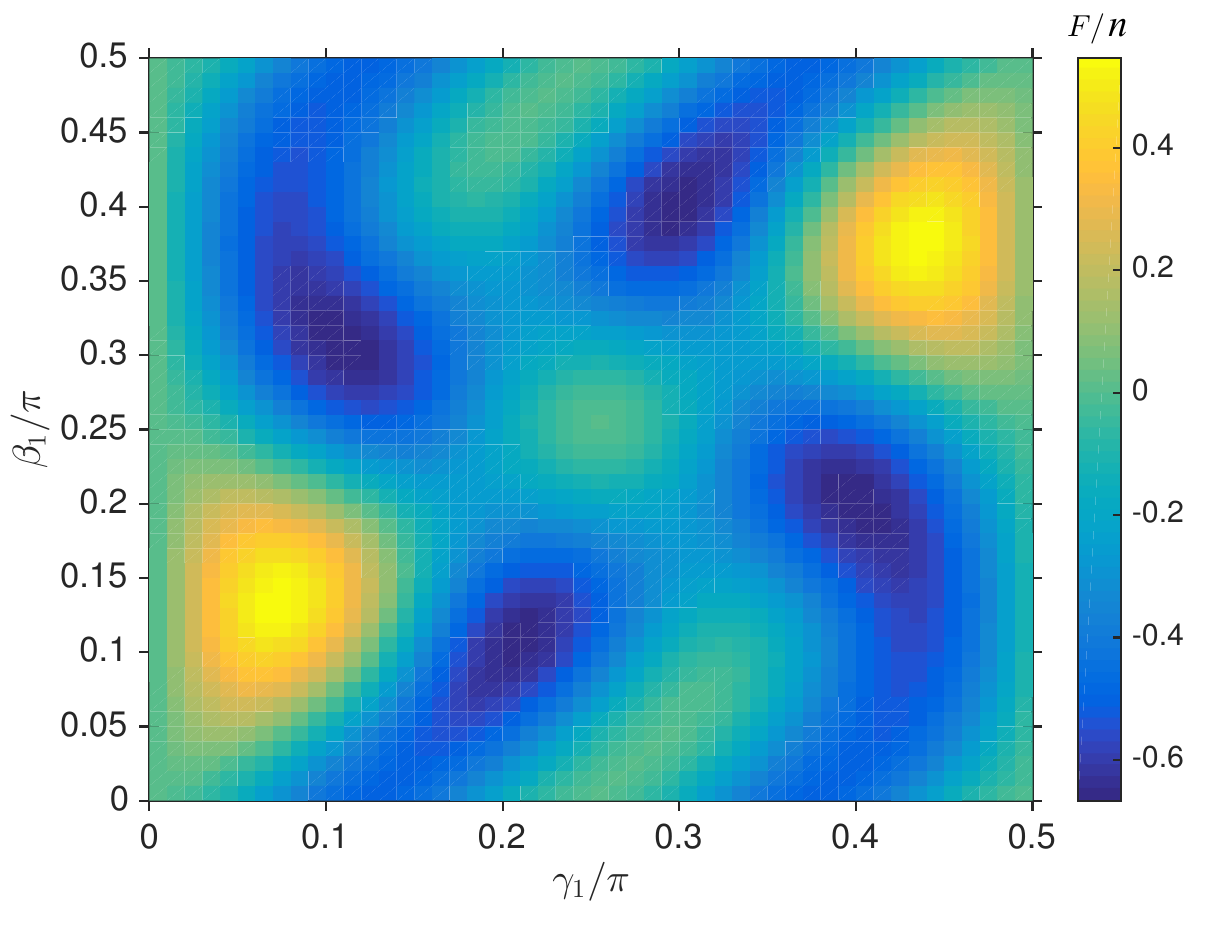}
		\caption{
The landscape of $F/n$ for $\QAOA_2$, in the sub-manifold
Eq.~(\ref{eq:manifold1}). The four darkest spots indicate the global minima
$F^*/n=-2/3$.  The origin $(0,~0)$ is a saddle point.  No local minima are
observed. The contour is symmetric w.r.t. $(\gamma_1,\beta_1)=(\pi/4,\pi/4)$,
reflecting the symmetry in Eq.~(\ref{eq:symmetry3}), (and the period $\pi/2$).
\label{fig:p2as}
			}  
	\end{center} 
\end{figure} 

 \subsection{Discussion: controllability and optimality}
\label{sec:landscape}

The optima of $F$ for QAOA level $p=1$ to 10 are tabulated in Appendix~\ref{sec:p2+}.
The optimal angles were obtained through numerical gradient descent search in
the sub-manifold Eq.~(\ref{eq:manifold1}).  The evaluation for $F_k$ in each
step is realized as Eq.~\eqref{eq:Fs2}, which only involves $2p$
multiplications of
2-by-2 matrices, and sum over $k$ gives $F$, so optimal angles for higher $p$
could be computed easily if desired.

Starting with a random initial guess of $(\bs\gamma,\bs\beta)$, the search
(with sufficiently small steps) always converges to a global minimum.
This behavior suggests
that at least within the sub-manifold, all local minima are global minima. 
For example,  for $p=2$,  there are two free parameters in the sub-manifold,
which we choose to be $\gamma_1$ and $\beta_1$.  Figure~\ref{fig:p2as} shows
the landscape of the expectation value $F$.  The four minima (darkest spots)
observed in one period ($\gamma_1,\beta_1,\in [0,\pi/2]$) are all global
minima.

This result calls for extended understanding of landscapes of quantum control.
In quantum control theory,  it has been shown that assuming
\emph{controllability}, i.e., evolution between any two states is achievable
via the set of controls given, the landscape of the infidelity $F$ over the
space of parameter values $(\bs\gamma, \bs\beta)$ generically has only global
minima~\cite{Rabitz04,Rabitz12,Rabitz16}.  Without controllability, the quantum
control landscape in general is rugged and admits local minima~\cite{Rabitz11}.

In the case of QAOA, the controls are constrained in a specific way: if
an infinite number of controls are allowed, i.e., $p\to \infty$, then the system
is controllable. The finite number of control steps dictated by the
level $p$ limits the controllability.  
For more general graphs, however, Eq.~\eqref{eq:th1} shows that for 
QAOA on MaxCut, 
even for $p=1$, there exists local optima in the space of parameter values.
For the special case of ring of disagrees, the system is still not controllable,
however, our numerical results indicate that, at least within the sub-manifold Eq.~(\ref{eq:manifold1}), all local minima are global minima.

 \section{Conclusions}
\label{sec:conclusion}
We studied parameter setting for QAOA on MaxCut. 
For $\QAOA_1$, we extended the results in
Ref.~\cite{Farhi2014}, providing an analytical expression for general graphs.
As a corollary, for triangle-free graphs with fixed vertex degree, 
the optimal angles for the driver Hamiltonian can be directly read off 
while the optimal angles
for the problem Hamiltonian show a dependence on the vertex degree.
For higher level $p$, direct analysis on the operator expansion becomes cumbersome, providing further
evidence that more advanced parameter setting techniques need to be 
developed.

For a special case of MaxCut, the ring of disagrees, 
which corresponds to a one-dimentional anti-ferromagnetic spin ring,
we analyze $\QAOA_p$, for arbitrary $p$,  
using a fermionic representation. 
Applying the Jordan-Wigner transformation 
transforms the evolution of an $\n$-qubit system 
under QAOA to a set of $\n/2$ independent evolutions within 
two-dimensional subspaces. The parameter setting problem, thus,
corresponds to finding optimal control parameters for an
ensemble of $\n/2$ non-interacting \pspins{} of spin-$\frac{1}{2}$.  
From this treatment we obtain
analytical expression for any $p$, and an easy numerical search for
the corresponding optimal angles.

The fermionic picture also enables us to explain symmetries in the
optimal parameters, suggesting a means to further reduce the effort
required to find optimal parameters by restricting to manifolds
defined by these symmetries.
The specific symmetry in the problem of ring of disagrees has its roots 
in the equal
footing of the action of the driver and the problem Hamiltonians -- each
corresponds to a single spin rotation.  
We observed numerically that within the parameter space, all minima are global 
minima.  While such a ``no-trap'' character of a quantum control landscape can 
be explained given controllability, the structure of QAOA for
finite $p$ often does not guarantee controllability. 
Future research that reveals the underlying theory may shed further light on 
the control landscape and 
the structure of QAOA, and inspire ways to simplify or improve the algorithm.

In Ref.~\cite{Farhi2014} it is conjectured that the best achievable approximation 
ratio for $\QAOA_p$ on a ring of size $n\ges 2p+2$ is $(2p+1)/(2p+2)$. 
The fermionic view we have presented, by simplifying the analysis, may be 
a useful step toward a
proof, but for the moment the conjecture remains open.  
Further work will examine how realistic noise affects the performance.  
While for certain simple noise models, including control noise affecting
the times for which the Hamiltonians are applied, can be analyzed within
the model, for other cases a combination of more sophisticated analytical
tools and experimentation on quantum hardware will be needed to evaluate
performance under noise. The simplicity of QAOA on the ring makes it a 
promising target for implementation on early quantum processors.

 \section{Acknowledgements}
We thank Sergey Knysh, Jason Dominy, Salvatore Mandr\`{a}, Bryan O'Gorman, Davide Venturelli,
Nick Rubin, Will Zeng and Robert Smith for enlightening discussions. 
S.H. was supported by NASA under award NNX12AK33A. He
would like to thank the Universities Space Research Association
(USRA) and the NASA Ames Research Center for the opportunity to participate in
the program that enabled this research.  
The authors would like to acknowledge support from the NASA Advanced
Exploration Systems program and the NASA Ames Research Center.
The views and conclusions contained herein
are those of the authors and should not be interpreted as necessarily
representing the official policies or endorsements, either expressed or
implied, of the U.S. Government.  
The U.S. Government is authorized to reproduce and distribute reprints for
Governmental purpose notwithstanding any copyright annotation thereon.
 
\appendix
\begin{appendices}
\section{Proof of Theorem 1}
\label{app:proof_maxcut}

\begin{proof}[Proof of Theorem 1]
For $p=1$, only terms corresponding to neighbors of $u$ and $v$ can contribute to the expectation of $C_{uv}$~\cite{Farhi2014}. 
We thus partition our objective function as
$$C= \frac12(I-\z_u\z_v) + C_u + C_v + \bar{C},$$
where $C_u$ is the $d$-many constraints involving only vertex $u$ but not $v$,
and $C_v$ is the $e$-many constraints involving only $v$. The remaining
constraints $\bar{C}$ do not contribute to the expectation value 
$\langle C_{uv} \rangle$. 
For simplicity, we write $(d,e,f)=(d_u, d_v,\lambda_{uv})$.

Let $c=\cos 2\beta$ and $s=\sin 2\beta$. We have 
\bqa
& & e^{i \beta B} \z_u\z_v e^{-i \beta B} \nnm
&=& c^2 \z_u\z_v + sc (\y_u\z_v + \z_u\y_v) + s^2 \y_u\y_v\;.
\eqa
The first term $\z_u\z_v$ commutes with $C$ and does not contribute to $\langle C_{uv}\rangle$.
 We conjugate each remaining term separately by $e^{i \gamma C}$. Let $c' = \cos \gamma$ and $s' = \sin \gamma$. We have
\bqa
& &\tr [{ \rho_0 e^{i\gamma C} \y_{u}\z_v e^{-i\gamma C} }] \nnm
&=& \tr\big[\rho_0 (I c' - i s' \z_u\z_v) \prod_{i=1}^d (I c' - i s' \z_u\z_{w_i})  \y_{u}\z_v \big]. \nnm
\eqa
Expanding the product on the right gives a sum of tensor products of Pauli operators. 
Clearly, the only term that can contribute is proportional to $\z_u\z_v * I^{\otimes d} * \y_u\z_v = -i \x_u$. 
Thus we have
\bqa
\tr\big[ \rho_0 e^{i\gamma C} \y_{u}\z_v e^{-i\gamma C}  \big]  = \tr\big[ \rho_0 (-i) s' c'^d (-i \x_u)  \big] =  - s' c'^d \nnm
 \eqa
By symmetry, we have
$
\tr[\rho_0 e^{i\gamma C} \z_{u}\y_v e^{-i\gamma C} \big]  =  - s' c'^e.
$
Observe that these terms are independent of the number of mutual neighbours (triangles) of $u$ and $v$. The next term is 
 \bqa
& & \tr\big[\rho_0 e^{i\gamma C} \y_{u}\y_v e^{-i\gamma C} \big]
=\tr\big[ e^{2i\gamma C_{u}} e^{2i\gamma C_{v}} \y_{u}\y_v \big] \nnm
& &= \tr\big[  \prod_{i=1}^d (c' I - i s' \z_u\z_{w_i}) \prod_{j=1}^e (c'I - i s' \z_v\z_{w_j})  \y_{u}\y_v   \big] \nnm
\eqa
The simplest term that contributes in this case is $\tr\big[\rho_0 f c'^{d+e-2} (-is')^2(-i \x_u) (-i\x_v)\big]= f c'^{d+e-2}  s'^2$.  
Corresponding to the triangles of $\langle{u}{v}\rangle$, in the above product we have $f$-many distinct values $i$ such that $w_i =w_j$.  As $\z_u\z_{w_i}*\z_u\z_{w_i}=I$, if $f>2$ then higher order terms depending on the number of triangles $f$ will contribute. For example, the next order terms will result from three pairs of $(\z_u\z_{w_i},\z_u\z_{w_i})$ and hence be proportional to  $s'^6$.
 Thus we have 
\bqa
\label{eqn:YYtermThm1proof}
& & \tr\big[\rho_0 (e^{i\gamma C} \y_{u}\y_v e^{-i\gamma C} ) \big] \nnm
 &=& \binom{f}{1}  c'^{d+e-2}  s'^2 +  \binom{f}{3}  c'^{d+e-6}  s'^6  \nnm
& & +  \binom{f}{5}  c'^{d+e-10}  s'^{10}+ \dots \nnm
 &=& c'^{d+e-2f} \sum_{i=1,3,5,\dots}^f \binom{f}{i}  (c'^{2})^{f-i}(s'^2)^i. 
\eqa
To sum this series, recall the binomial theorem, which we may split into even and odd sums as 
\bqa
& &  \sum_{i=0,2,\dots}^f \binom{f}{i} a^{f-i} b^{i} + \sum_{i=1,3,\dots}^f \binom{f}{i} a^{f-i} b^{i}  \nnm
& =& \sum_{i=0}^f \binom{f}{i} a^{f-i} b^{i}=(a+b)^f
\eqa
which also gives
\bqa
& &  \sum_{i=0,2,\dots}^f \binom{f}{i} a^{f-i} b^{i} - \sum_{i=1,3,\dots}^f \binom{f}{i} a^{f-i} b^{i} \nnm
& &=  \sum_{i=0}^f (-1)^i \binom{f}{i} a^{f-i} b^{i} = (a-b)^f,
\eqa
and hence 
\bqa
  \sum_{i=1,3,\dots}^f \binom{f}{i} a^{f-i} b^{i} = \frac12 ( (a+b)^f - (a-b)^f) \;.
\eqa
Thus the above sum becomes 
\bqa
& &  \sum_{i=1,3,\dots}^f \binom{f}{i}  (c'^{2})^{f-i}(s'^2)^i 
= \frac12 (1 - \cos^f 2\gamma)
\eqa
which yields
\begin{equation}  \label{eqn:YYtermSumThm1proof}
 \tr\big[\rho_0 (e^{i\gamma C} \y_{u}\y_v e^{i\gamma C} ) \big]  = \frac12  c'^{d+e-2f} (1 - \cos^f 2\gamma)
\end{equation}
Putting this all together, we have
\bqa \label{eq:A11}
\langle C_{uv}\rangle &=& \tr\big[\rho_0 e^{i\gamma C} e^{i \beta B} C_{uv} e^{-i \beta B}  e^{-i\gamma C}  \big] \nnm
&=& \frac12 - \frac{sc}{2} \tr\big[\rho_0 e^{i\gamma C} (\y_u\z_v + \z_u\y_v) e^{-i\gamma C} ) \big] \nnm 
& &- \frac{s^2}{2} \tr\big[\rho_0 e^{i\gamma C} \y_u\y_v e^{-i\gamma C} ) \big] \nnm \nnm
&=&  \frac12 + \frac12 s c s' (c'^d + c'^e) -\frac14 s^2c'^{d+e-2f}(1-\cos^f 2\gamma). \nnm
\eqa
\end{proof}

\section{Proof of symmetry relation Eq.~(\ref{eq:symmetry}) }
\label{sec:app1}
We prove Eq.~(\ref{eq:symmetry})
\bqa
F_k(\bd \gamma,\bd\beta)=F_k(-\bd\beta',-\bd\gamma') 
\eqa
where 
\bqa
\bd\gamma \eq (\gamma_1,\gamma_2,\ldots, \gamma_{p-1},\gamma_p) \\
\bd\beta \eq (\beta_1,\beta_2,\ldots, \beta_{p-1},\beta_p) \\
\bd\gamma' \eq (\gamma_p,\gamma_{p-1},\ldots, \gamma_2, \gamma_1) \\
\bd\beta' \eq (\beta_p,\beta_{p-1},\ldots, \beta_2, \beta_1)\;. 
\eqa

\begin{proof}
We consider a unitary operator
$
R = \cos\frac{\theta}{2} \z + \sin\frac{\theta}{2} \x
$
which rotates a Bloch vector about axis $(\hat k + \hat z)$ by $\pi$.
Note that $R^\dag=R$, $R^2= 1$ and
\bqa
R\z R \eq \hat k \cdot  \hat \sigma \nnm
R \hat k \cdot  \hat \sigma R \eq \z \nnm
R U_B(\beta) R \eq U_C(\beta) \nnm
R U_C(\gamma) R \eq U_B(\gamma)\;,
\eqa
we have
\bqa
RUR \eq R U_B(\beta_p) RR U_C(\gamma_p) R \ldots R U_B(\beta_1) RR U_C(\gamma_1) R \nnm
\eq   U_C(\beta_p)  U_B(\gamma_p)  \cdots  U_C(\beta_1)  U_B(\gamma_1)  \nnm
\eq   \big[U_B(-\gamma_1) U_C(-\beta_1) \cdots  U_B(-\gamma_p)  U_C(-\beta_p)      \big]^\dag \nnm
\eq U'^\dag
\eqa
where $U'\equiv U(-\bd\beta',-\bd\gamma')$.

Insert $R^2=\bd 1$ to $F_k$ we get
\bqa
F_k(\bd \gamma,\bd\beta) \eq \tr [(\hat k \cdot  \hat \sigma)  U  \z  U^\dag] \nnm
\eq \tr [R(\hat k \cdot  \hat \sigma)  RR U RR \z  RR U^\dag R] \nnm
\eq \tr [\z RUR (\hat k \cdot  \hat \sigma) RU^\dag R ] \nnm
\eq \tr [\z U'^\dag (\hat k \cdot  \hat \sigma) U'] \nnm
\eq \tr [(\hat k \cdot  \hat \sigma) U' \z U'^\dag ] \nnm
\eq F_k(-\bd\beta',-\bd\gamma') 
\eqa
\end{proof}
 \section{Detailed results for \texorpdfstring{$p=2$}{p=2} and higher}
\label{sec:p2+}

\begin{table*} [!htbp] \begin{center}
\begin{tabular}{ccccccccccccccccccccccccc}
\textbf{$p  $}&\textbf{$r$}&\textbf{$  F^*/n $}&\textbf{$ \gamma_1  $}&\textbf{$  \beta_1 $}&\textbf{$   \gamma_2 $}&\textbf{$ \beta_2  $}&\textbf{$\gamma_3 $}&\textbf{$ \beta_3 $}&\textbf{$\gamma_4 $}&\textbf{$ \beta_4  $}&\textbf{$ \gamma_5 $}&\textbf{$ \beta_5 $}\\\hline
\hline
1&3/4&-1/2&0.1250&&&&&&&&&\\\hline
2&5/6&-2/3&0.2052&0.1026&&&&&&&&\\\hline
3&7/8&-3/4&0.2268&0.1888&0.0918&&&&&&&\\\hline
4&9/10&-4/5&0.2357&0.2161&0.1791&0.0850&&&&&&\\\hline
5&11/12&-5/6&0.2403&0.2282&0.2094&0.1724&0.0802&&&&&\\\hline
6&13/14&-6/7&0.3035&0.1639&0.2506&0.2835&0.0794&0.2409&&&&\\\hline
7&15/16&-7/8&0.2303&0.1623&0.3468&0.2690&0.1042&0.2397&0.1599&&&\\\hline
8&17/18&-8/9&0.2445&0.1638&0.2839&0.3484&0.1539&0.1530&0.2581&0.1291&&\\\hline
9&19/20&-9/10&0.1929&0.1648&0.3307&0.3016&0.1551&0.2538&0.2174&0.1089&0.3117&\\\hline
10&21/22&-10/11&0.2208&0.1374&0.3098&0.2974&0.2702&0.1205&0.3148&0.1904&0.1423&0.2572\\\hline
\end{tabular}
 \caption{Optimal angles for different levels of QAOA. Angles are in unit $\pi$. Gradient descend search is implemented with the optimal angles for level $p$ set to be the initial guess for level $p+1$. Arbitrary initial guess also always converges to a global minimum of $F(\bs \gamma,\bs\beta)$ in the sub-manifold Eq.~(\ref{eq:manifold1}). Multiple sets of optimal angles exist for $p\ges 2$, only one of them is shown for each level.
\label{tb:optangle}
}
\end{center}
\end{table*}

\begin{figure*} [htbp]
	\begin{center}  
				\includegraphics[width=4in]{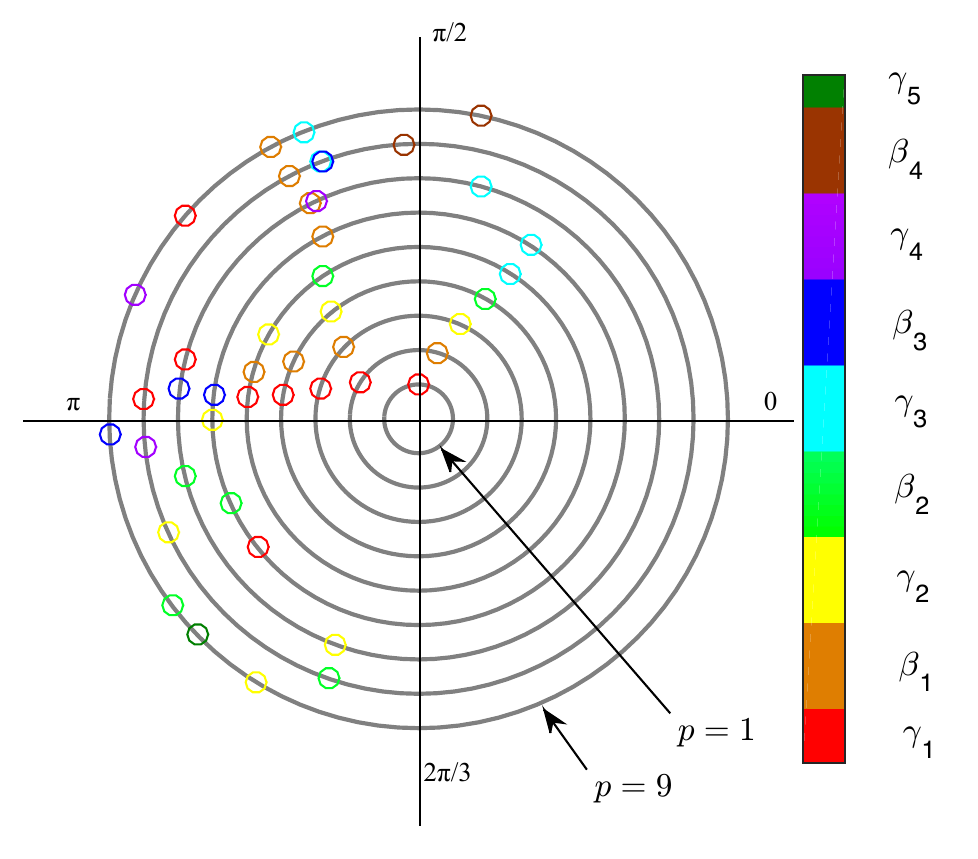}
		\caption{
		Optimal angles in the submanifold defined by Eq.~\eqref{eq:manifold1}. The optimal points are plotted on the complex plane with angles $(4\bs\gamma,4\bs\beta)$ as the argument and radius given by the level, $p=1$ to $p=9$ from the inner to outer circles. 
			}  
	\label{fig:opt_angle_circle}
	\end{center} 
\end{figure*}

For $p=2$,
terms in that are non-vanishing to $F$ is

\begin{widetext}
\bqa
\frac{F}{n} 
\eq \frac{1}{64} 
\big[
-7\cos(4\beta_1 + 4\beta_2 + 4\gamma_1 + 4\gamma_2) - 6\cos(4\beta_1 + 4\beta_2 + 4\gamma_1)  \nnm
& & + 3\cos(4\beta_1 + 4\beta_2 - 4\gamma_1 + 4\gamma_2) + 4\cos(4\beta_1 + 4\beta_2 +  4\gamma_2) \nnm
& & + 3\cos(4\beta_1 - 4\beta_2 + 4\gamma_1 + 4\gamma_2) - 6\cos(4\beta_1 - 4\beta_2 + 4\gamma_1) - \nnm
& & 3\cos(4\beta_1 - 4\beta_2 - 4\gamma_1 + 4\gamma_2) + 4\cos(4\beta_1 + 4\gamma_1 + 4\gamma_2) - \nnm
& & 4\cos(4\beta_1 + 4\gamma_1) - 4\cos(4\beta_1 + 4\gamma_2) - 3\cos(-4\beta_1 + 4\beta_2 + 4\gamma_1 + 4\gamma_2) \nnm
& & + 6\cos(-4\beta_1 + 4\beta_2 + 4\gamma_1) + 3\cos(-4\beta_1 + 4\beta_2 - 4\gamma_1 + 4\gamma_2) + \nnm
& & 7\cos(-4\beta_1 - 4\beta_2 + 4\gamma_1 + 4\gamma_2) + 6\cos(-4\beta_1 - 4\beta_2 + 4\gamma_1) - \nnm
& & 3\cos(-4\beta_1 - 4\beta_2 - 4\gamma_1 + 4\gamma_2) - 4\cos(-4\beta_1 - 4\beta_2 + 4\gamma_2) - \nnm
& & 4\cos(-4\beta_1 + 4\gamma_1 + 4\gamma_2) + 4\cos(-4\beta_1 + 4\gamma_1) + 4\cos(-4\beta_1 + 4\gamma_2) - \nnm
& & 6\cos(4\beta_2 + 4\gamma_1 + 4\gamma_2) - 6\cos(4\beta_2 - 4\gamma_1 + 4\gamma_2) - 4\cos(4\beta_2 + 4\gamma_2) \nnm
& & + 6\cos(-4\beta_2 + 4\gamma_1 + 4\gamma_2) + 6\cos(-4\beta_2 - 4\gamma_1 + 4\gamma_2)  \nnm
& & + 4\cos(-4\beta_2 + 4\gamma_2) 
\big]\;.
\eqa
\end{widetext}

If limited in the sub-manifold Eq.~(\ref{eq:manifold1}),
\begin{widetext}
\bqa
\frac{F}{n}
\eq \frac{1}{64} \Big(
    -2\cos(8\beta_1) + 3\cos(8\beta_1 + 8\gamma_1) - 12\cos(4\beta_1 + 8\gamma_1) \nnm
& & -8\cos(4\beta_1 + 4\gamma_1) + 12\cos(4\beta_1 - 8\gamma_1) + 8\cos(4\beta_1-4\gamma_1) \nnm
& & + 7\cos(8\beta_1- 8\gamma_1) - 8\cos(8\beta_1 - 4\gamma_1) + 6\cos(8\gamma_1) \nnm
& & + 8\cos(4\gamma_1) - 14
\Big)\;.
\eqa
\end{widetext}

In Table.~\ref{tb:optangle} we show numerical for optimal angles for higher QAOA levels in the manifold Eq.~\eqref{eq:manifold1} (multiple optima were found for $p\ges 2$, we show only one for each $p$).  
The same sets of angles are also plotted on the circles in Fig.~\ref{fig:opt_angle_circle}.

 \end{appendices}

\newpage
\bibliographystyle{unsrt}
\bibliography{bibQAOA}

\begin{thebibliography}{10}

\bibitem{Farhi2014}
Edward Farhi, Jeffrey Goldstone, and Sam Gutmann.
\newblock A {Quantum} {Approximate} {Optimization} {Algorithm}.
\newblock {\em arXiv:1411.4028}, November 2014.

\bibitem{Farhi2014b}
Edward Farhi, Jeffrey Goldstone, and Sam Gutmann.
\newblock A {Quantum} {Approximate} {Optimization} {Algorithm} {Applied} to a
  {Bounded} {Occurrence} {Constraint} {Problem}.
\newblock {\em arXiv:1412.6062}, December 2014.

\bibitem{Farhi2016}
Edward Farhi and Aram~W. Harrow.
\newblock Quantum {Supremacy} through the {Quantum} {Approximate}
  {Optimization} {Algorithm}.
\newblock {\em arXiv:1602.07674}, February 2016.

\bibitem{Shabani16}
Zhi-Cheng Yang, Armin Rahmani, Alireza Shabani, Hartmut Neven, and Claudio
  Chamon.
\newblock Optimizing variational quantum algorithms using Pontryagin's minimum
  principle.
\newblock{\em Phys. Rev. X}, 7(2):021027, 2017.

\bibitem{Jiang17}
Zhang Jiang, Eleanor Rieffel, and Zhihui Wang.
\newblock Near-optimal quantum circuit for Grover's unstructured search using a transverse field
\newblock {\em Phys. Rev. A}, 95(6):062317, 2017.

\bibitem{Hadfield17}
Stuart Hadfield, Zhihui Wang, Bryan O’Gorman, Eleanor G. Rieffel, Davide
Venturelli, and Rupak Biswas. 
\newblock  From the quantum approximate optimization algorithm to a quantum alternating operator ansatz.
\newblock {\em Algorithms} 12 (2), 34, 2019.

\bibitem{barak2015beating}
Boaz Barak, Ankur Moitra, Ryan O'Donnell, Prasad Raghavendra, Oded Regev, David
  Steurer, Luca Trevisan, Aravindan Vijayaraghavan, David Witmer, and John
  Wright.
\newblock Beating the random assignment on constraint satisfaction problems of
  bounded degree.
\newblock {\em arXiv:1505.03424}, 2015.

\bibitem{wecker2016training}
Dave Wecker, Matthew~B. Hastings, and Matthias Troyer.
\newblock Training a quantum optimizer.
\newblock {\em Phys. Rev. A}, 94(2):022309, 2016.

\bibitem{preskill_quantum_2012}
John Preskill.
\newblock Quantum computing and the entanglement frontier.
\newblock {\em arXiv:1203.5813}, March 2012.

\bibitem{boixo_characterizing_2016}
Sergio Boixo, Sergei~V. Isakov, Vadim~N. Smelyanskiy, Ryan Babbush, Nan Ding,
  Zhang Jiang, John~M. Martinis, and Hartmut Neven.
\newblock Characterizing {Quantum} {Supremacy} in {Near}-{Term} {Devices}.
\newblock {\em arXiv:1608.00263}, July 2016.

\bibitem{lieb_two_1961}
Elliott Lieb, Theodore Schultz, and Daniel Mattis.
\newblock Two soluble models of an antiferromagnetic chain.
\newblock {\em Annals of Physics}, 16(3):407--466, December 1961.

\bibitem{barouch_statistical_1970}
Eytan Barouch, Barry~M. McCoy, and Max Dresden.
\newblock Statistical {Mechanics} of the {XY} {Model}. {I}.
\newblock {\em Phys. Rev. A}, 2(3):1075--1092, September 1970.

\bibitem{Rabitz04}
Herschel~A. Rabitz, Michael~M. Hsieh, and Carey~M. Rosenthal.
\newblock Quantum optimally controlled transition landscapes.
\newblock {\em Science}, 303(5666):1998--2001, 2004.

\bibitem{Rabitz12}
Re-Bing Wu, Ruixing Long, Jason Dominy, Tak-San Ho, and Herschel Rabitz.
\newblock Singularities of quantum control landscapes.
\newblock {\em Phys. Rev. A}, 86:013405, Jul 2012.

\bibitem{Rabitz16}
Benjamin Russell, Herschel Rabitz, and Rebing Wu.
\newblock Quantum control landscapes are almost always trap free.
\newblock {\em arXiv:1608.06198}, 2016.

\bibitem{Rabitz11}
Re-Bing Wu, Michael~A. Hsieh, and Herschel Rabitz.
\newblock Role of controllability in optimizing quantum dynamics.
\newblock {\em Phys. Rev. A}, 83:062306, Jun 2011.

\end{thebibliography}

\noindent Correction: A missing factor of 2 in Eq. (\ref{eq:th1}) has been inserted
and a sign error in Eq. (\ref{eq:A11}) has been fixed.

\end{document}